\documentclass[aps,prx,twocolumn,floatfix,superscriptaddress]{revtex4}

\usepackage{amsmath}
\usepackage{amssymb}
\usepackage{nicefrac}
\usepackage{braket}
\usepackage[pdftex]{graphicx}
\usepackage{txfonts}
\usepackage{color}

\renewcommand{\vec}[1]{\boldsymbol{#1}}

\begin{document}

\title{The family of topological Hall effects for electrons in skyrmion crystals}

\author{B{\"o}rge G{\"o}bel}
\email[]{bgoebel@mpi-halle.mpg.de}
\affiliation{Max-Planck-Institut f\"ur Mikrostrukturphysik, D-06120 Halle (Saale), Germany}

\author{Alexander Mook}
\affiliation{Institut f\"ur Physik, Martin-Luther-Universit\"at Halle-Wittenberg, D-06099 Halle (Saale), Germany}

\author{J\"urgen Henk}
\affiliation{Institut f\"ur Physik, Martin-Luther-Universit\"at Halle-Wittenberg, D-06099 Halle (Saale), Germany}

\author{Ingrid Mertig}
\affiliation{Max-Planck-Institut f\"ur Mikrostrukturphysik, D-06120 Halle (Saale), Germany}
\affiliation{Institut f\"ur Physik, Martin-Luther-Universit\"at Halle-Wittenberg, D-06099 Halle (Saale), Germany}

\date{\today}

\begin{abstract}
Hall effects of electrons can be produced by an external magnetic field, spin orbit-coupling or a topologically non-trivial spin texture. The topological Hall effect (THE) --- caused by the latter --- is commonly observed in magnetic skyrmion crystals. Here, we show analogies of the THE to the conventional Hall effect (HE), the anomalous Hall effect (AHE), and the spin Hall effect (SHE). In the limit of strong coupling between conduction electron spins and the local magnetic texture the THE can be described by means of a fictitious, `emergent' magnetic field. In this sense the THE can be mapped onto the HE caused by an external magnetic field. Due to complete alignment of electron spin and magnetic texture, the transverse charge conductivity is linked to a transverse spin conductivity. They are disconnected for weak coupling of electron spin and magnetic texture; the THE is then related to the AHE\@. The topological equivalent to the SHE can be found in antiferromagnetic skyrmion crystals. We substantiate our claims by calculations of the edge states for a finite sample. These states reveal in which situation the topological analogue to a quantized HE, quantized AHE, and quantized SHE can be found.
\end{abstract}
    
\maketitle

\section{I\lowercase{ntroduction}} 
If an electric current is applied to an electrical conductor in the presence of a static magnetic field in perpendicular direction, a voltage across the sample is observed. This `Hall effect' of electrons was discovered almost 140 years ago~\cite{hall1879new}. In the following and especially in the last 20 years it has been found that the external magnetic field is not mandatory to produce a Hall effect and can be exchanged for example by spin-orbit coupling~\cite{nagaosa2010anomalous} or a chiral magnetic texture~\cite{bruno2004topological} that produces an effective magnetic field, that is a `Berry curvature'~\cite{berry1984quantal}. The family of Hall effects for electrons in solids was extended by the quantum Hall effect~\cite{landau1930diamagnetismus,hofstadter1976energy,klitzing1980new}, anomalous Hall effect~\cite{nagaosa2010anomalous}, quantum anomalous Hall effect~\cite{liu2008quantum,chang2013experimental}, spin Hall effect~\cite{d1971possibility,kato2004observation}, quantum spin Hall effect~\cite{kane2005quantum,bernevig2006quantum}, and topological Hall effect~\cite{neubauer2009topological,schulz2012emergent,kanazawa2011large, lee2009unusual,li2013robust,bruno2004topological,hamamoto2015quantized, yin2015topological,lado2015quantum,gobel2017THEskyrmion, gobel2017QHE,ndiaye2017topological}.

The Hall resistivity
\begin{align}
\rho_{xy} = \rho_{xy}^{\mathrm{HE}}+\rho_{xy}^{\mathrm{AHE}}+\rho_{xy}^{\mathrm{THE}}
\end{align}
is the off-diagonal element of the resistivity tensor $\rho$, that connects electric field $\vec{E}$ and current density $\vec{J}$ via Ohm's law $\vec{E} = \rho\vec{J}$. It is given by contributions from the ordinary Hall effect (HE), the anomalous Hall effect (AHE) and eventually the topological Hall effect (THE) in magnetic systems. The HE can appear quantized as quantum Hall effect (QHE) and is due to an external magnetic field that couples to the charge of the conduction electrons; the electrons are deflected by a Lorentz force in a semiclassical picture. The AHE arises due to spin-orbit coupling (SOC) with intrinsic (Berry curvature) and extrinsic (skew-scattering and side jump) contributions~\cite{nagaosa2010anomalous}, all of which deflect electrons with opposite spin into opposite directions. In a ferromagnet, the imbalance of majority and minority conduction electrons leads to both spin and charge transport. In non-magnetic materials the charge accumulation at both sides of the sample is compensated, leading to a pure spin Hall effect (SHE).

This Paper addresses the THE of electrons in topologically non-trivial magnetic textures $\vec{s}(\vec{r})$ in two dimensions. The probably most prominent example is the skyrmion~\cite{skyrme1962unified,bogdanov1989thermodynamically, bogdanov1994thermodynamically,rossler2006spontaneous,muhlbauer2009skyrmion}. Its winding is quantified by the topological charge density
\begin{align}
	n_\mathrm{Sk} (\vec{r})  =  \vec{s}(\vec{r}) \cdot \left[ \frac{\partial \vec{s}(\vec{r})}{\partial x}  \times  \frac{\partial \vec{s}(\vec{r})}{\partial y}  \right]
\end{align}
that integrates to the integer topological charge (or skyrmion number)~\cite{nagaosa2013topological}
\begin{align}
	N_\mathrm{Sk}  =  \frac{1}{4\pi} \int_{xy} n_\mathrm{Sk}(\vec{r})\, \mathrm{d}^{2} r.
\end{align}
The spins of the conduction electrons tend to align with the magnetic texture, thereby accumulating a Berry phase~\cite{everschor2014real}. If skyrmions are arranged periodically in a skyrmion crystal (SkX) this Berry phase can be translated to a Berry curvature in reciprocal space that deflects traversing electrons and leads to nonzero transverse Hall coefficients for charge and spin.
Skyrmion crystals have first been observed in non-centrosymmetric B20 bulk-compounds like MnSi~\cite{muhlbauer2009skyrmion} and FeGe~\cite{yu2011near}. They also appear at interfaces, like in Fe/Ir~\cite{heinze2011spontaneous}.

\begin{figure*}
  \centering
  \includegraphics[width = \textwidth]{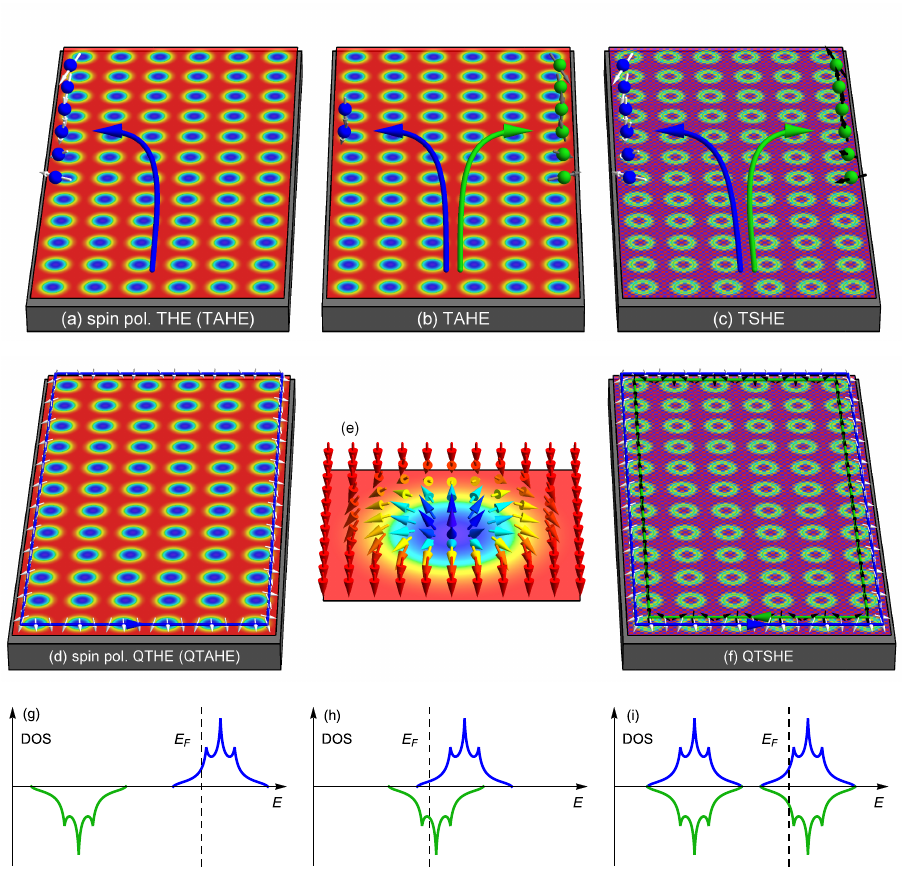}
  \caption{Variety of topological Hall effects. (a) In the strong-coupling limit the electron spin aligns with the skyrmion texture and exhibits a topological version of the spin-polarized HE\@. Only one spin-type of carrier contributes to transport at a certain energy, and spin and charge conductivities are inseparable. This makes the spin-polarized THE a special case of a TAHE\@. (b) In the weak-coupling limit the local spin polarization is incomplete, leading to an unequal number of spin parallel and antiparallel electrons being deflected into opposite directions. This gives nonzero, uncoupled spin and charge conductivities, i.\,e., a TAHE\@. (c) In an antiferromagnetic skyrmion crystal an equal number of spin parallel electrons on the two sublattices are deflected into opposite directions leading to zero charge conductivity. Since the sublattice textures are oriented oppositely this leads to a pure topological spin Hall effect. The large arrows represent the motion of the conduction electrons as response to an applied electric field in longitudinal direction. Electrons are deflected in transverse direction due to interaction with the texture. They accumulate charge (spheres) and spin represented by small arrows (white: parallel; black: antiparallel and gray levels: partially oriented spins). The colormap represents the $z$ orientation of the magnetic texture. (e) Shows the unit cell of the skyrmion crystal for $\lambda=10$. An arrow represents the magnetic moment $\vec{s}_i$ at lattice site $i$. (d) and (f) are the quantized versions of the topological Hall effects from (a)--(c), respectively. They are observable if the Fermi energy is located in a band gap between two Landau levels. Edge states are represented by the colored lines, where the color represents the spin channel. In the case of spin mixing no quantized version of the TAHE is observed. (g)--(i) Density of states of the zero-field band structure (schematic) and position of the Fermi energy for the corresponding scenarios.}
  \label{fig:overview}
\end{figure*}

This Paper is dedicated to characterize precisely the THE and to update the classification of electronic Hall effects. The THE is commonly mentioned in line with the other Hall effects. We think it is more sensible to speak of topological versions of the conventional, the anomalous and the spin Hall effect because the THE manifests itself differently in dependence on the strength of the Zeeman interaction of electron spin and magnetic texture. 
%In the strong-coupling limit the THE transforms into the topological version of a locally fully-spin-polarized Hall effect that also appears quantized. 
We suggest a nomenclature (Fig.~\ref{fig:overview}) for the variety of THEs and tell which effect can be expected in the different scenarios.

In the strong-coupling limit, the Hamiltonians of THE and a spin-polarized QHE can be mapped onto each other (Sect.~\ref{sec:QTHE}). This locally fully spin-polarized quantized topological Hall effect (QTHE) is the extreme case of a quantized topological anomalous Hall effect (QTAHE)\@. In the weak-coupling limit spin and charge transport coefficients are decoupled, what corresponds to the typical anomalous Hall scenario in its topological version (TAHE; Sect.~\ref{sec:TAHE}). The topological analogue to the spin Hall effect~\cite{gobel2017afmskx,buhl2017topological,akosa2017theory} (TSHE), i.\,e., pure spin currents, is found in antiferromagnetic skyrmion crystals and completes the topological Hall trio (Sect.~\ref{sec:TSHE}; see Fig.~\ref{fig:overview}). By calculating edge states of a finite sample we illustrate when these topological Hall effects show up in their quantized version.

\section{M\lowercase{odel and methods}}
In this Paper we focus on the topological contribution to the Hall effect that arises due to the nonzero topological charge density $n_\mathrm{Sk}(\vec{r_i})$ of the magnetic texture formed by localized magnetic moments $\{\vec{s_i}\}$. Therefore, we consider the tight-binding Hamiltonian
\begin{align} 
  H &  =  \sum_{\braket{ij}} t \,c_{i}^\dagger c_{j} + m \sum_{i} \vec{s}_{i} \cdot (c_{i}^\dagger \vec{\sigma}c_{i})
  \label{eq:ham_the} 
\end{align} 
that features nearest-neighbor hopping of conduction electrons ($c_{i}^\dagger$ creation and $c_{i}$ annihilation operators at site $i$, amplitude $t$), and Hund's coupling (quantified by $m$), i.\,e., the Zeeman interaction of the electrons' spins with the fixed magnetic texture $\{ \vec{s}_i \}$ [visualized in Fig.~\ref{fig:overview}(e)]. The magnetic texture is thought to be produced by localized electrons that are not explicitly featured in the calculation.

To segregate the effect of the magnetic texture, we neglect (i) SOC which would contribute to the AHE and (ii) the influence of external magnetic fields that are commonly needed to stabilize a SkX phase; they would contribute to the conventional or quantized Hall effect. 

Having diagonalized the Hamiltonian, the eigenenergies $E_n(\vec{k})$ and the eigenvectors $\ket{u_n(\vec{k})}$, both depending on the wave vector $\vec{k}$, allow to calculate the Berry curvature~\cite{berry1984quantal}
\begin{align}
  \Omega_n^{(xy)}(\vec{k}) &  = -2\,\mathrm{Im} \sum_{m \ne n}
  \frac{\braket{u_{n\vec{k}} | \partial_{k_x} H_{\vec{k}}| u_{m\vec{k}}}\braket{u_{m\vec{k}}|\partial_{k_y} H_{\vec{k}} | u_{n\vec{k}}}}{(E_{n\vec{k}} - E_{m \vec{k}})^2},
\end{align}
whose integral over the Brillouin zone (BZ) yields the Chern number
\begin{align}
	C_{n}  =  \frac{1}{2\pi} \int_{\mathrm{BZ}}\Omega_{n}^{(xy)}(\vec{k})\, \mathrm{d}^{2}k
\end{align}
of band $n$ and the topological Hall conductivity~\cite{nagaosa2010anomalous}
\begin{align}
  \sigma_{xy}(E_\mathrm{F}) &  =  -\frac{e^{2}}{h} \frac{1}{2\pi} \sum_{n} \int_{\mathrm{BZ}} \Omega_{n}^{(xy)}(\vec{k}) \, f(E_{n}(\vec{k}) - E_\mathrm{F}) \,\mathrm{d}^{2}k. \label{eq:cond}
\end{align}

Concerning spin transport in a ferromagnet, one velocity $v_x = \partial_{k_x} H_{\vec{k}}$ has to be replaced by the spin current operator $J_x = \frac{1}{2}\{v_x,\mathcal{M}\}$, with $\mathcal{M} = \operatorname{diag}(\sigma_z,\dots,\sigma_z)$ that accounts for spin-up and spin-down orientation with respect to the global quantization axis $\vec{M}  =  M\,\vec{e}_z$~\cite{gradhand2012first}. The noncollinear texture of a skyrmion requires to consider the local quantization axis (at each site) by means of the operator $\mathcal{M}  =  \operatorname{diag}(\vec{s}_1 \cdot \vec{\sigma},\dots,\vec{s}_n \cdot \vec{\sigma})$. We obtain the spin Berry curvature
\begin{align}
\Omega_{n,xy}^\mathrm{S}(\vec{k})& = -2\,\mathrm{Im} \sum_{m \ne n}
  \frac{\braket{u_{n\vec{k}} | J_x | u_{m\vec{k}}} \braket{u_{m\vec{k}}|\partial_{k_y} H_{\vec{k}} | u_{n\vec{k}}}}{(E_{n\vec{k}} - E_{m \vec{k}})^2}\label{eq:spinberry}
\end{align}
and the spin Hall conductivity for spin polarization parallel to the local quantization axes
\begin{align}
\sigma_{xy}^\mathrm{S}(E_\mathrm{F})&  =  \frac{e}{4\pi} \frac{1}{2\pi}\sum_{n} \int_{\mathrm{BZ}} \Omega_{n,xy}^\mathrm{S}(\vec{k}) \, f(E_{n}(\vec{k}) - E_\mathrm{F}) \,\mathrm{d}^{2}k. \label{eq:condspin}
\end{align}

\section{I\lowercase{nfluence of the} H\lowercase{und's coupling on the band structure}}
Before presenting results for the transverse charge and spin conductivities, the effect of the Hund's coupling on the electron spin and on the band structure is prerequisite. The strength of the interaction  is quantified by $\nicefrac{m}{t}$.

Aiming at the simplest, yet most distinct manifestation of topological properties of the SkX in band structure and transport coefficients, we consider only s-electrons on a square structural lattice (lattice constant $a$). The respective zero-field band structure (band structure of the structural lattice for $m  =  0$) has one band,
\begin{align}
E_0(\vec{k}) = 2t\left[\cos(k_xa)+\cos(k_ya)\right],
\end{align}
with a maximum at $4t$, a minimum at $-4t$, and two saddle points at $E_\mathrm{VHS}  =  0$ (van Hove singularity, VHS). On top of this, it is spin-degenerate.

The square magnetic unit cell of the SkX, with a period of $\lambda \cdot a$, leads to back-folding of the bands into the magnetic Brillouin zone and thus to $n  =  \lambda^2$ bands. A nonzero coupling $m$ breaks time reversal symmetry, and the spin degeneracy is lifted. This is accompanied by a shift of 
the bands to lower (spin antiparallel to the magnetic texture) and higher (spin parallel to the magnetic texture) energies. For $m > 4t$ (half the band width of $E_0$), the band structure is separated into two blocks, the electron spin is almost completely aligned with the magnetic texture, and small band gaps show up. These findings are best understood in the strong-coupling limit $\nicefrac{m}{t}\rightarrow \infty$.

\section{Q\lowercase{uantized topological} H\lowercase{all effect in the strong-coupling limit}} \label{sec:QTHE}

\subsection{Transformation to the emergent field}
In the limit $\nicefrac{m}{t}\rightarrow\infty$ the electron spin completely aligns with the texture to minimize the system's energy. In this case the Zeeman term of the Hamiltonian can be diagonalized by applying a set $\{ U_i \}$ of unitary transformations,
\begin{align}
U_i^\dagger (\vec{s}_i\cdot\vec{\sigma}) U_i &  =  \sigma_z,  \label{eq:rho}
\end{align}
so that the spin at site $i$ points into the $z$ direction of the local coordinate system. The new electron operators read $\tilde{c}_i = U_i^\dagger c_i$, while the new hopping strength $\tilde{t}_{ij}  =  U_i^\dagger t U_j$ is now a non-diagonal (2$\times$2) matrix. The full transformation is discussed in Ref.~\onlinecite{hamamoto2015quantized}. 

In the strong-coupling limit, spin-parallel electrons cannot align antiparallel to the magnetic texture and vice versa. For the parallel aligned electrons, only the $(1,1)$ element 
\begin{align}
\tilde{t}_{ij}^{(1,1)} = t\left(\cos\frac{\theta_i}{2}\cos\frac{\theta_j}{2}+\sin\frac{\theta_i}{2}\sin\frac{\theta_j}{2}\exp[-\mathrm{i}(\phi_i-\phi_j)]\right)
\end{align}
of the hopping is relevant, which depends on the azimuthal $\theta_i$ and the polar angle $\phi_i$ of the local magnetic moment at site $i$. It can be transformed into an effective hopping strength
\begin{align}
t_{ij}^\mathrm{(eff)} = t \,\mathrm{e}^{\mathrm{i} a_{ij}} \cos\frac{\theta_{ij}}{2},
 \label{eq:effectivehopping}
\end{align}
with $\cos\theta_{ij}  =  \vec{s}_i\cdot\vec{s}_j$ the angle between the moments at sites $i$ and $j$; the phase is given by 
\begin{align}
a_{ij} = \arctan\frac{-\sin(\phi_i-\phi_j)}{\cos(\phi_i-\phi_j)+\cot\frac{\theta_i}{2}\cot\frac{\theta_j}{2}}.
\end{align}
With these substitutions the Hamiltonian becomes
\begin{align}
H = \sum_{\braket{ij}}t\cos\frac{\theta_{ij}}{2}\,\exp{(\mathrm{i} a_{ij})}\,d_i^\dagger d_j,
\end{align}
which describes fully spin-aligned electrons (equivalent to spinless electrons; creation $d_i^\dagger$ and annihilation $d_i$). The constant Zeeman term has been dropped. The cosine in \eqref{eq:effectivehopping} merely scales the hopping strength and converges to $1$ for large skyrmions ($\lambda\rightarrow\infty$); the THE has then been mapped onto the QHE described by the Hamiltonian 
\begin{align}
H = \sum_{\braket{ij}}t\,\exp(\mathrm{i} b_{ij})\,d_i^\dagger d_j.
\end{align}
In the quantum Hall effect spinless electrons interact with a uniform magnetic field $\vec{B}$, that is oriented out of plane (spin-polarized electrons can be described likewise). This field couples to the charge, rather than to the spin of the conduction electrons and affects their kinetic energy. In the tight-binding description this leads to complex hopping amplitudes with phase factors
\begin{align}
  b_{ij} = e / \hbar \int_{\vec{r}_{i} \to \vec{r}_{j}} \vec{A}(\vec{r}) \cdot \mathrm{d}\vec{l},
\end{align}
whose phase is determined by the vector potential $\vec{A}$, with $\vec{B}=\nabla\times\vec{A}$. The above integral is along the hopping path.

%Keep in mind, that our description of of the THE only considers coupling of the magnetic texture to the electrons' spin. Coupling to the charge is not considered. It would lead to a conventional (or quantized) Hall effect covered in $\rho_{xy}^\mathrm{HE}$ in Eq.~\eqref{eq:rho}. 

The phase $a_{ij}$ for the topological Hall effect can be identified with the phase $b_{ij}$ from the quantum Hall effect. The local skyrmion density $n_\mathrm{Sk}(\vec{r})$ can therefore be identified with a fictitious magnetic field that is collinear but inhomogeneous.

Application of the unitary transformations $\{U_i\}$ is exact; however, taking only a diagonal element is strictly valid only in the strong-coupling limit $\nicefrac{m}{t}\rightarrow \infty$. In this limit the Hund's coupling can be expressed as an interaction of the skyrmion's emergent field with the charge of the conduction electrons. Considering the $(2,2)$ element instead of the $(1,1)$ element, one would have described the lower block of bands, in which the electron spin is aligned antiparallel to the magnetic texture. In that case, the emergent field as well as the topological Hall conductivity change sign.

Describing the QHE on a lattice restricts the magnetic field to discrete values, so that the magnetic flux per magnetic unit cell is an integer multiple of the flux quantum $\Phi_0  =  \nicefrac{h}{e}$. Therefore, a large magnetic unit cell has to be considered to account for the appropriate phase $b_{ij}$. The magnetic flux per plaquette then reads $\Phi = \nicefrac{p}{q} \,\Phi_0$; $p$ and $q$ are coprime integers~\cite{hofstadter1976energy}. The emergent field of the skyrmion induces an integer number of flux quanta per unit cell as well. Both systems differ by the fact that the emergent field is not homogeneous; still one can find a quantum Hall system with a magnetic field equal to the average emergent field of the SkX. The average flux per plaquette then reads $\Phi = \nicefrac{N_\mathrm{Sk}}{n} \,\Phi_0 = \nicefrac{p}{q} \,\Phi_0$.

This equivalence of THE and QHE in the limit $\nicefrac{m}{t}\rightarrow\infty$ allows to carry over all the known results from one effect to the other. Bands in the THE can, for example, be interpreted as dispersive Landau levels which carry a Chern number of $+\mathrm{sign}(N_\mathrm{Sk})$ [$-\mathrm{sign}(N_\mathrm{Sk})$] for the upper [lower] block. One result is a quantized transverse conductivity, provided the Fermi level lies within a band gap. Furthermore, $\sigma_{xy}$ behaves peculiarly for Fermi energies near VHSs, as is discussed in the next section.

\subsection{Conductivity}
The analogy of THE and QHE holds strictly speaking in the limit $\nicefrac{m}{t} \rightarrow \infty$. Nevertheless, the emergent-field description is reasonable as long as block separation in the band structure is given ($m \ge 4t$). We emphasize that the computations were performed with the exact Hamiltonian~\eqref{eq:ham_the}; the numerical results are only interpreted by means of the emergent field.

\begin{figure}
  \centering
  \includegraphics[width = 0.9\columnwidth]{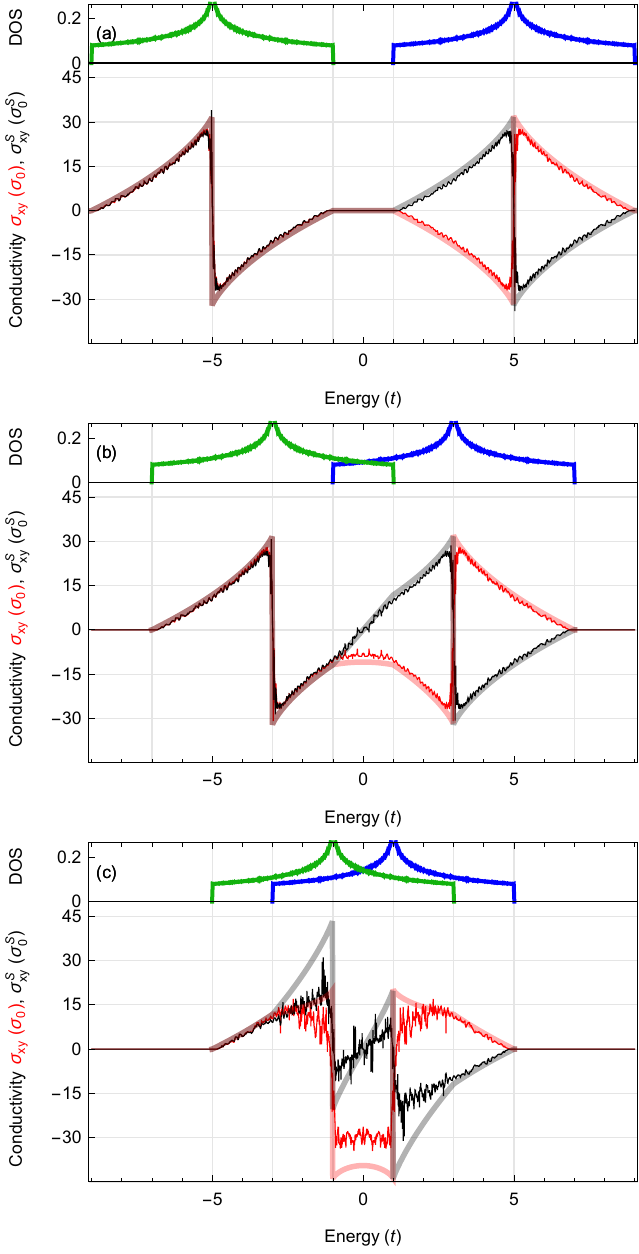}    
  \caption{Charge and spin Hall conductivity for different $\nicefrac{m}{t}$ regimes. (a) $\nicefrac{m}{t} = 5$, (b) $\nicefrac{m}{t} = 3$ and (c) $\nicefrac{m}{t} = 1$. Red represents the charge conductivity in units of $\sigma_0 = e^2/h$, black is the spin Hall conductivity in units of $\sigma_0^\mathrm{S} = e/(4\pi )$. The upper panel shows the density of states (DOS) of the zero-field band shifted by $\pm m$. Additive and subtractive superposition of the integrated DOS (fermion character is respected by sign) yields the transparent red and black curves, respectively. They approximate the corresponding conductivities well, if the spin is aligned with the texture. For small $m$ it still captures qualitative features. The skyrmion size is $\lambda=8$.}
  \label{fig:cond}
\end{figure}

The transverse conductivity [red in Fig.~\ref{fig:cond}(a)] depends strongly on the position of the Fermi energy. The two blocks in the band structure, easily recognized in the spectrum, show opposite sign, which is explained by the opposite spin alignment with respect to the magnetic texture, giving emergent fields of opposite sign. This leads to deflection of electrons into opposite transverse directions in a semi-classical picture. It is sufficient to discuss only the upper block.

Interpreting the bands as dispersive Landau levels with a Chern number of $+1$ tells that the conductivity decreases stepwise upon increasing $E_\mathrm{F}$. Due to the inhomogeneity of the emergent field, the bands are `deformed' and have a nonzero band width. The spectrum consists therefore of energy regions in which $\sigma_{xy}$ is quantized [insulating system; QTHE; Fig.~\ref{fig:overview}(d)] or not quantized [metallic system; THE; Fig.~\ref{fig:overview}(a)].

The energies of the Landau levels are determined by the constraint that, like for the QHE, for each Landau level the number of states is identical (Onsager's quantization scheme). At these energies, the corresponding Fermi line of the zero-field band structure encloses an area of 
\begin{align}
\zeta_i = \zeta_0 \left( i-\frac{1}{2} \right), \quad i = 1,2, \ldots, n,
\end{align}
in the Brillouin zone; $\zeta_0$ is the area of the (smaller) magnetic Brillouin zone. The curvature of these  Fermi lines dictates both the fermion character and the sign of the transverse conductivity: electron-like (hole-like) orbits give a negative (positive) sign for energies below (above) $E_\mathrm{VHS}$. The sign of the transverse conductivity changes at the VHS~\cite{gobel2017THEskyrmion,gobel2017QHE,gobel2018magnetoelectric,hatsugai2006topological,arai2009quantum}, because it separates electron- from hole-like Fermi lines. The particular Landau level closest to the VHS thus carries a large Chern number of $1-n$ which compensates the Chern number of all other bands (for details see Ref.~\onlinecite{gobel2017QHE}). The zero-field band structure dictates therefore the shape of $\sigma_{xy}(E_\mathrm{F})$: different lattices lead to different spectra [cf.\ a triangular lattice in Ref.~\onlinecite{gobel2017THEskyrmion} and a honeycomb lattice in Ref.~\onlinecite{gobel2017QHE}].

Based on the above observations we developed an easy way to predict $\sigma_{xy}(E_\mathrm{F})$ (see Ref.~\onlinecite{gobel2017QHE}): the DOS is integrated up to the Fermi energy and shifted at VHSs, the result is the transparent red curve in the background of Fig.~\ref{fig:cond}(a).

\subsection{Edge states and higher order skyrmions}
The bulk-boundary correspondence connects the sum of Chern numbers of the occupied bands with the number of topologically protected edge states~\cite{Hatsugai1993,Hatsugai1993a}, the latter being computed by Green function renormalization for the semi-infinite system~\cite{henk1993subroutine,bodicker1994interface}.

\begin{figure*}
  \centering
  \includegraphics[width = 0.95\textwidth]{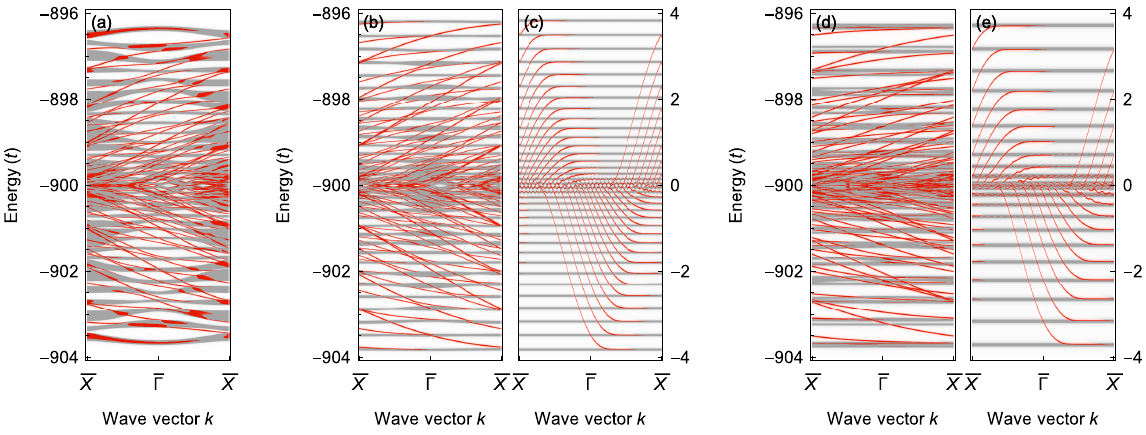}
  \caption{Edge states of skyrmion crystals. (a) Skyrmion crystal and (b) manipulated SkX in the the strong-coupling limit $\nicefrac{m}{t} = 900$ and for $\nicefrac{N_\mathrm{Sk}}{n}  =  \nicefrac{1}{36}$. (c) Quantum Hall system for $\nicefrac{p}{q}  =  \nicefrac{1}{36}$. (d) SkX with $\nicefrac{N_\mathrm{Sk}}{n}  =  \nicefrac{3}{64}$ and (e) quantum Hall system with $\nicefrac{p}{q} = \nicefrac{3}{64}$. Gray: bulk states; red: edge states.}
  \label{fig:surfaceTHEvsQHE}
\end{figure*}

Figure~\ref{fig:surfaceTHEvsQHE}(a) illustrates the edge states of a skyrmion crystal in the strong-coupling limit $\nicefrac{m}{t} =  900$. Due to the large degree of spin polarization we show only the lower block. The number of edge states changes by $1$ for high and low energies. The edge states change from left to right propagation at $E_\mathrm{VHS}$, which complies with the sign change of the conductivity.

In Fig.~\ref{fig:surfaceTHEvsQHE}(b) the system has been manipulated by renormalizing the hopping [$t\rightarrow t/\cos(\theta_{ij}/2)$ according to Eq.~\eqref{eq:effectivehopping}]. Additionally, the skyrmion texture has been tilted so that the emergent field becomes almost homogeneous. This way, the topological charge of a skyrmion remains invariant but the SkX mimics a quantum Hall system with dispersionless Landau levels [cf.\ Figs.~\ref{fig:surfaceTHEvsQHE}(b) and (c)] even better. 
The dispersion relations of edge states in the quantum Hall system and the SkX are very similar, especially their slope. Keep in mind that the SkX magnetic unit cell comprises $\lambda \times \lambda$ sites, while for the QH system it comprises $n \times 1$ sites (with $n=\lambda^2$ equal to the number of atoms in the magnetic unit cell). The $\Gamma$X direction of the QH system is $\lambda=6$ times as long as that of the SkX.

The equivalence of the spin polarized QTHE and the QHE holds also for larger skyrmion numbers. The ratio $\nicefrac{N_\mathrm{Sk}}{n}$ dictates the ratio $\nicefrac{p}{q}$ for the related quantum Hall system. Figs.~\ref{fig:surfaceTHEvsQHE}(d) and (e) show the results for the ratio $\nicefrac{3}{64}$, i.\,e., for a skyrmion with $N_\mathrm{Sk}=3$. The expansion 
\begin{align}
\frac{3}{64} = \frac{1}{21+\frac{1}{3}}
\end{align}
tells that the band structure shows $21$ bundles that are connected by surface states like in the case $\nicefrac{p}{q}  =  \nicefrac{1}{21}$ (Refs.~\onlinecite{chang1995berry,chang1996berry}). Each of these bundles consists of three subbands with a total Chern number of $-1$, except for the bundle at the VHS, which consists of four bands with a total Chern number of $+20$, compensating the other Chern numbers. This behavior is well understood for the QHE and can be carried over to the topological version. The Diophantine equation~\cite{dana1985quantised,thouless1982quantized,chang1996berry} tells the Chern numbers of the subbands. Furthermore, the essential physics are represented by Hofstadter butterflies~\cite{hofstadter1976energy}; such figures depict the band energies versus magnetic flux, which is characterized by the ratio $\nicefrac{p}{q}$ for the QH case and can now be understood as the magnetic flux of the emergent field given by $\Phi=\nicefrac{N_\mathrm{Sk}}{n}\,\Phi_0$.

\subsection{Spin conductivity}
The transverse spin conductivity of the SkX in the strong-coupling limit is shown in Fig.~\ref{fig:cond} (panel a; black curve). Charge and spin transport appear inseparably linked, which is attributed to the high degree of spin polarization. In the case of non-overlapping blocks ($m\ge 4t$) the results are well resembled by the emergent field picture. The spin conductivity is given by $\sigma_{xy}^\mathrm{S}  =  -\nicefrac{\hbar}{2e} \,\sigma_{xy}$ for all Fermi energies in the upper block. For the lower block the charge conductivity is sign-reversed but the spin conductivity is not. Due to the opposite sign of the emergent field, electrons with opposite spins are deflected into opposite directions. 

In the strong-coupling limit, the THE becomes a spin-polarized quantized topological Hall effect, which is  related to a spin-polarized QHE brought about by a magnetic field [Fig.~\ref{fig:overview}(a)]. The spin-polarized QHE is also the extreme case of a QAHE. Therefore the spin-polarized THE can also be labeled TAHE [Fig.~\ref{fig:overview}(a)] and the quantized version, for Fermi energies located in a band gap, can also be labeled QTAHE [Fig.~\ref{fig:overview}(d)]. The Hall conductivity and the transverse spin conductivity of both compared systems agree well.

Note, that we calculate a spin conductivity in a local coordinate system. The measured spin accumulation (in a global system) is determined by this quantity but the texture at the sample's edge has to be taken into account [cf. accumulated spins in Fig.~\ref{fig:overview}(a)].

Commonly, in an AHE scenario  charge and spin transport are (strictly) uncoupled. This happens in the weak-coupling limit, in which complete spin-polarization
is not given for all Fermi energies.

\section{T\lowercase{opological anomalous} H\lowercase{all effect in the weak-coupling limit}} \label{sec:TAHE}
In materials with weaker Hund's coupling $m$, the separation into non-overlapping blocks of bands is not given, hence the emergent-field picture becomes invalid.

In collinear ferromagnets the Hall effect can be treated well in a two-channel model, where each channel deals with one kind of either spin parallel or spin antiparallel electrons. This picture breaks down if the two spin species hybridize, for example, due to spin-orbit coupling or, as in our case, due to a noncollinear magnetic texture. When the electron spin does not follow the texture adiabatically, complete local spin polarization, i.\,e., parallel or antiparallel alignment, is not given. Still, we utilize the two channel model and compare its prediction to the actual results.

Figures~\ref{fig:cond}(b) and (c) show charge and spin conductivities for $\nicefrac{m}{t} = 3$ and $\nicefrac{m}{t} = 1$, respectively. Both conductivities change sign at the energy of the zero-field VHS $E_\mathrm{VHS} = \pm m$. For energies, at which the two zero-field bands do not overlap ($|E|>4t-m$) both curves are well explained by the two-channel model, because the electron spin is almost completely aligned with the magnetic texture. 

For energies at which the blocks overlap, the two-channel picture describes qualitatively the spectrum but fails to predict the amplitude, especially for small $\nicefrac{m}{t}$. This is due to the fact that the spin is locally not aligned parallel or antiparallel to the texture. The bands become dispersive, and the eigenstates of the Hamiltonian show contributions from both spin species.

For $m<4t$ the charge conductivity can exceed the spin Hall conductivity in magnitude (in units of their respective quanta) if spin-antiparallel hole-like states hybridize with spin-parallel electron-like states for energies $|E_\mathrm{F}|<2t-|m-2t|$ [cf.\ $|E_\mathrm{F}| < t$ for $\nicefrac{m}{t} = 1$ and $\nicefrac{m}{t}  =  3$ in Figs.~\ref{fig:cond}(b) and (c)]. In the two-channel approximation, then, electrons with opposite spins are deflected into the same direction and the spin conductivity can drop to zero.

For $m<2t$ the spin conductivity can exceed the charge conductivity. This can happen for $-4t+m<E_\mathrm{F}<-m$, where spin-parallel electron-like states mix with spin-antiparallel electron-like states, and for $m<E_\mathrm{F}<4t-m$, where spin-parallel hole-like states mix with spin-antiparallel hole-like states. In the two-channel approximation, electrons with opposite spins are deflected into opposite directions [this corresponds to the scenario depicted in Fig.~\ref{fig:overview}(b)]. This increases the spin conductivity and decreases the charge conductivity compared to the case of non-overlapping zero-field bands. Due to the missing alignment the actual transverse conductivities are always reduced in magnitude compared to the results expected for the two-channel model.

The scenario of mixed zero-field blocks corresponds to the AHE in the stricter sense [cf. Fig.~\ref{fig:overview}(b)], i.\,e., where both types of spin carriers dominate the transport effects. In this case the emergent field interpretation breaks down and bands do not behave like Landau levels. They become dispersive and band gaps disappear. Therefore, edge states are visible but are always superposed by bulk bands and the quantized version of the TAHE for mixed carriers cannot be observed. 

For Fermi energies $|E_\mathrm{F}|>4t-m$, where only one spin species is present, the local spin alignment is given and bands become merely flat. This QTAHE is again equivalent to a spin-polarized version of the QHE as in the strong-coupling limit (cf. Sect.~\ref{sec:QTHE}).

\section{T\lowercase{opological spin} H\lowercase{all effect in antiferromagnetic skyrmion crystals}} \label{sec:TSHE}
The nontopological AHE in a ferromagnet arises due to SOC and a net magnetization. In a SkX, the effect of the SOC is taken over by the magnetic texture, which also induces the block separation. Therefore, a pure topological spin Hall effect --- the number of spin-parallel electrons is identical to the number of spin-antiparallel electrons at $E_\mathrm{F}$ --- could only be achieved when the texture is removed. Consequently, all transverse charge and spin transport phenomena would vanish as well.

The topological analogue of the SHE~\cite{buhl2017topological,gobel2017afmskx,akosa2017theory} has been established in antiferromagnetic (AFM) skyrmions~\cite{zhang2016antiferromagnetic,barker2016static,zhang2016magnetic} and antiferromagnetic SkXs~\cite{gobel2017afmskx}. These textures consist of two sublattices or layers with conventional skyrmions, but with opposite magnetic moments in the two sublattices. Hence, the topological charge and the emergent field of each sublattice have opposite signs, leading to a zero total topological charge and a vanishing THE\@.

In this section we briefly review the results of Ref.~\onlinecite{gobel2017afmskx} to complete the family of topological Hall effects. First, we consider only hopping within a sublattice. The AFM-SkX then consists of two non-interacting SkXs on the two sublattices. The conduction electrons are deflected into opposite directions depending on the sublattice on which they `live' [Fig.~\ref{fig:overview}(c)]. Parallel spins (arbitrarily labeled as `up') in one sublattice are interpreted as antiparallel spins in the other sublattice (also `up'). For the calculation of the spin Hall conductivity this is respected by a modified $\mathcal{M}  =  \operatorname{diag}(\vec{s}_1^A \cdot \vec{\sigma},-\vec{s}_1^B \cdot \vec{\sigma},\dots,\vec{s}_n^A \cdot \vec{\sigma},-\vec{s}_n^B \cdot \vec{\sigma})$ that enters Eq.~\eqref{eq:spinberry} via $J_x$. 

The results can be understood as two copies of Fig.~\ref{fig:overview}(a). Locally spin-parallel electrons from sublattice A (spin up) are deflected to the left, and parallel electrons from sublattice B (spin down because of the reversed magnetization) are deflected to the right. This leads to a zero charge Hall conductivity and a nonzero spin Hall conductivity, i.\,e., a TSHE\@. $\sigma_{xy}^\mathrm{S}(E_\mathrm{F})$ looks just like for the SkX, but with twice the magnitude because of the two sublattices. Within the band gaps, the TSHE is quantized [QTSHE, Fig.~\ref{fig:overview}(f)].

In real materials, nearest-neighbor hopping, i.\,e., inter-sublattice hopping, is allowed. Nevertheless, the THE in an AFM-SkX vanishes, in contrast to a topological spin Hall effect, as shown in Ref.~\onlinecite{gobel2017afmskx}. $\sigma_{xy}^\mathrm{S}(E_\mathrm{F})$ appears modified by the altered zero-field band structure.

\section{C\lowercase{onclusion}}
We have analyzed the electron transport in different coupling regimes in conventional and antiferromagnetic skyrmion crystals. 

In the strong-coupling limit, the topological equivalent to a fully spin-polarized QHE is found and explained in the emergent-field picture. Because the electron spin follows the texture adiabatically, charge and spin transport are twined and it can therefore also be seen as a special case of a TAHE or a QTAHE\@.

In the weak-coupling limit, spin-mixing terms in the Hamiltonian prevent the interpretation by means of the emergent field picture. The charge conductivity is no longer quantized and the spin conductivity is not proportional to the charge conductivity. Here, one is confronted with a topological analogue to the typical anomalous Hall effect: an unequal number of electrons with their spin aligned parallel and antiparallel to the magnetic texture propagate into opposite transverse directions, resulting in a decoupling of charge and spin conductivities.

An extreme version of the anomalous Hall effect is the spin Hall effect, with identical numbers of electrons with spin parallel and antiparallel to the magnetic texture. This effect is not possible in a skyrmion crystal, but in an antiferromagnetic skyrmion crystal. We find a TSHE or QTSHE, depending on the position of the Fermi energy.

We have shown that topologically nontrivial magnetic textures lead to a topological Hall effect, which is viewed not as a separate Hall effect but as the topological version of well-known Hall effects previously discussed in ferromagnets or conventional metals. 

\begin{acknowledgments}
This work is supported by SPP 1666 and SFB 762 of Deutsche Forschungsgemeinschaft (DFG).\\ \vspace{\baselineskip}
\end{acknowledgments}

\section*{Author contribution}
B. G{\"o}bel conducted all calculations. All authors discussed the results and contributed to the manuscript.

\bibliography{short,MyLibrary}
\bibliographystyle{apsrev}

\end{document}